\documentclass[aps,prd, ,preprintnumbers, nofootinbib,twocolumn]{revtex4-1}
\usepackage{bm}
\usepackage{verbatim}
\usepackage{latexsym}
\usepackage{amsmath,amsfonts,amssymb}
\usepackage{graphicx,epsfig}
\usepackage{psfrag}
\usepackage{amsthm}
\usepackage{color}
\usepackage{braket}
\usepackage{multirow} 

\usepackage{amsmath}
\usepackage{braket}
\usepackage{indent first}
\usepackage[normalem]{ulem}
\useunder{\uline}{\ul}{}

\begin{document}
\title{Reply to ``Comment on `Quantum Raychaudhuri Equation' ''}
\author{Saurya Das}
\email{saurya.das@uleth.ca}
\affiliation{Theoretical Physics Group and Quantum Alberta, Department of Physics and Astronomy, University of Lethbridge, 4401 University Drive, Lethbridge, Alberta T1K 3M4, Canada}

\begin{abstract}

The above comment \cite{ld}
claims that the paper ``Quantum Raychaudhuri Equation''
by S. Das, Phys. Rev. {\bf D89} (2014) 084068
has ``problematic points'' with regards to its derivation and implications.
We show below that the above claim is incorrect, and that there are no problems with results of \cite{1}
or its implications.

\end{abstract}

\maketitle

The starting {\bf assumptions} of ``Quantum Raychaudhuri Equation''
by S. Das, Phys. Rev. {\bf D89} (2014) 084068 (ref.\cite{1} hereafter) were: \\
(i) The background spacetime is classical satisfying Einstein equations with suitable boundary conditions (page 1, left column of ref.\cite{1}).\\
(ii) Classical geodesics are replaced by quantal trajectories, as is natural in a quantum (as opposed to a classical) Universe. As explained in \cite{1},
the Bohmian formulation of quantum mechanics is completely equivalent to the standard formulation of quantum mechanics. The quantal trajectories are defined using the wavefunction and correctly predicts the behavior of all observed quantum phenomena. Therefore, no experiments or observations can invalidate the Bohmian picture \cite{2,3}.

\noindent
In deriving the relativistic Quantum Raychaudhuri Equation (QRE), \cite{1} also assumes:\\
(iii) The Klein-Gordon or scalar field equation in the
{\it geometrical optics (eikonal) approximation} describes relativistic quantum particles. The standard covariant definitions of momentum and velocity for the field, which are used to define quantal trajectories, have nothing to do with the Bohmian formalism {\it per se} (Eqs.(10-13) of Ref.\cite{1}).
This procedure has been studied extensively in the context of gravitational lensing (refs.\cite{4,5} below and refs.[16,17] of ref.[1]).

\vspace{0.2cm}
Next, quantal trajectories have the following well-known {\bf properties}:\\
(a) Two such trajectories do not meet or cross \cite{2,3,4,5,6,7,8}.
Hence they do not form conjugate points in a given manifold.
This is by virtue of the fact that quantal trajectories are governed by first order differential equations (such as Eqs.(6,12) of \cite{1}. \\
(b) physical properties associated with these trajectories (such as position, momentum, energy and signature) are not measured at any intermediate stage (as any such measurement would result in the collapse of the wavefunction and of the trajectories), and these can assume any values. Contrary to the authors of the comment, there is nothing ill-defined about this, and it is in fact these deviations from classical values in the intermediate stages, a consequence of the ``quantum potential'' (Eqs.(8),(9),(16),(17) of \cite{1}),
that give rise to new quantum phenomena. This bears similarity to properties of internal loops in quantum field theory.

\vspace{0.2cm}
The {\bf main results} of \cite{1} follow directly from properties (a) and (b) above, namely: \\
1. The Hawking-Penrose singularity theorems do not apply to these trajectories, and singularities in the sense of these
theorems do not exist (Section II of \cite{1}).
This is because the existence of conjugate points is an essential ingredient in the proofs of these theorems \cite{9}.\\
2. Extreme curvature regions are inaccessible to quantal trajectories (Section III of \cite{1}).

Note that results 1 and 2 above do not depend on the precise form of the quantum correction terms in the QRE (Eq.(17) of \cite{1}).

\vspace{0.1cm}
Ref.\cite{1} does not claim one of the following:\\
I. Quantal trajectories `remove' spacetime singularities. The latter are merely inaccessible to them.\\
II. A full theory of quantum gravity is not required. On the contrary, it acknowledges that for such a theory the smooth manifold structure may break down at small scales (concluding section of \cite{1}).

The comment in question \cite{ld}
is written with the same motivation as \cite{1},
that of replacing the classical Raychaudhuri equation with a quantum version. It also starts with assumptions (i), (ii) and (iii) above and practically follows all intermediate steps of \cite{1}, including the Bohmian formulation of quantum mechanics.
Their Eqs.(1-5) (equation numbers of the comment in question \cite{ld},
refer to a significantly shortened version of the original comment, arXiv:1606.04738,
published in Phys. Rev. D)
are identical to equations (10, 12, 15-17) of \cite{1},
and they arrive at their version of QRE [Eq.(12)] with one extra term. The same can be said about their geodesic deviation equation (13), when compared to Eq.(18) of \cite{1}.
Note that the explicit form of the induced metric was not used in
\cite{1} to derive its main results,
namely 1 and 2 above, or equivalently those following Eqs.(17) and (18) of \cite{1}.
As a result, Eq.(17) of \cite{1} contains the induced metric $h^{ab}$.
Using an explicit form (Eq.(7) in comment \cite{ld} taking into account the magnitude of the velocity field)
results in an additional term in the quantum Raychaudhuri equation.
However, we remind the reader, and as explained in property (b) above,
no measurements are done at intermediate points, and hence these trajectories are not required to remain timelike at such points.
Second and more importantly as remarked earlier,
the main results of \cite{1} {\it do not} depend on the precise form of the quantum corrections.
They simply follow from property (a) and (b) above.
In other words, regardless of the precise form of the QRE,
the results 1 and 2 above would remain unchanged.
It follows that further implications of these results are correct as well
\cite{10,11,12}.

Finally, the authors claim on p.4 of \cite{ld}
that the quantal trajectories do not form a congruence in the presence of gravity.  This is incorrect. Quantal (Bohmian) trajectories are governed by first order equations even in the presence of gravity (Eq.(13) of \cite{1}), and therefore form a congruence.

In summary, the authors of the comment \cite{ld}
have misunderstood the derivation of the Quantum Raychaudhuri equation and its implications for the singularity theorems, and their claims are flawed. Much of this was already explained in \cite{13}.

%

\end{document}